\documentclass[10pt,pra,superscriptaddress,amsmath,amssymb,twocolumn]{revtex4-1}
\pdfoutput=1

\usepackage{amsthm}
\usepackage{graphicx}
\usepackage[all]{xy}

\newcommand{\bra}[1]{\langle #1 |}
\newcommand{\ket}[1]{| #1 \rangle}

\newcommand{\ketbra}[2]{\ket{#1}\bra{#2}}
\newcommand{\proj}[1]{\ket{#1}\bra{#1}}
\newcommand{\tr}{{\rm Tr}\,}
\newcommand{\one}{{\bf 1}}
\newcommand{\re}{{\rm Re}\,}

\newcommand{\tortau}{t}

\newcommand{\mysection}[1]{\section{#1}}

\newcommand{\chan}[1]{\mathcal #1}

\newcommand{\id}{{\rm id}}

\newcommand{\imag}{{\rm Im}}
\newcommand{\real}{{\rm Re}}

\theoremstyle{plain}

\theoremstyle{definition}

\begin{document}

\title{Perturbative quantum error correction}
\author{C\'edric B\'eny}
\affiliation{Centre for Quantum Technologies, National University of Singapore, 3 Science Drive 2, Singapore 117543}
\affiliation{Institut f\"ur Theoretische Physik, Leibniz Universit\"at Hannover Appelstra\ss e 2, 30167 Hannover, Germany}
\date{February 18, 2011}

\begin{abstract}
We derive simple necessary and sufficient conditions under which a quantum channel obtained from an arbitrary perturbation from the identity can be reversed on a given code to the lowest order in fidelity. We find the usual Knill-Laflamme conditions applied to a certain operator subspace which, for a generic perturbation, is generated by the Lindblad operators. For a weak interaction with an environment, the error space to be corrected is a subspace of that spanned by the interaction operators, selected by the environment's initial state. 
\end{abstract} 

\maketitle

The ability to fully control physical systems so as to faithfully encode or transmit quantum information is of great interest both for fundamental research and applications. The main obstacle is decoherence, a form of noise which rapidly destroys the quantum nature of a state. Quantum error correction (QEC) techniques consist in encoding quantum information into a physical system in such a way that it can be actively saved from such effects. This requires certain assumptions about the noise, one of which is generally that it is weak in some sense.
 In the most common framework, one assumes that arbitrary errors can happen independently on each qubit with a small probability, so that the likelihood of a combination of $n$ errors happening at the same time goes as the $n$th power of that small probability. This intuition was generalized in Ref.~\cite{knill00} to any type of perturbative noise generated by a weak Hamiltonian interaction with an environment, or by a Markovian evolution. 
Nevertheless, our general understanding of the correctability of a code still relies on the Knill-Laflamme (KL) conditions~\cite{knill97} which have been proven to be necessary and sufficient only when applied to the Kraus operators of an exactly known channel, and not in a perturbative framework. 

This problem was considered in Ref.~\cite{leung97} where sufficient conditions for perturbative QEC to lowest order in a small parameter were derived. Ref.~\cite{knill00} considered simpler sufficient conditions which amount to applying the KL conditions to a certain set of operators. Here we show that the KL conditions are necessary and sufficient when applied to a subset of these operators.

Specifically, we assume that the noise is modeled by a quantum channel $\chan N (\rho)=\sum_i N_i \rho N_i$ whose Kraus operators $N_i$ are convergent power series in a small parameter $\epsilon$, and such that $\chan N(\rho) = \rho$ for $\epsilon = 0$. Without loss of generality we write
\begin{equation}
\label{exp}
N_i = \delta_{0i} \one + \epsilon E_i + \epsilon^2 F_i + O(\epsilon^3).
\end{equation}
Our result states that correction to lowest significant order in $\epsilon$ is possible if and only if the operators $\one,E_1,\dots,E_m$, and hence their linear span, satisfy the KL conditions (Equ.~\ref{zecond}). Note that this excludes $E_0$. This operator must be anti-hermitian for the channel to be trace-preserving, and can be eliminated to order $\epsilon^2$ by the unitary $e^{-\epsilon E_0}$.

This result applies to any one-parameter family of channels expressed themselves (instead of their Kraus operators) as a power series in the parameter $t$. For the generic case, the first term to be corrected is of order $t$ and has the Lindblad form. Our result says that the KL conditions must be satisfied for all Lindblad operators and the identity (Equ.~\ref{linbdlequ}). 

We also study the one-parameter family of channels resulting from a constant Hamiltonian interaction with an environment. In this case, the first nontrivial term to be corrected appears to order $t^2$. It is sufficient for the span of the interaction operators to satisfy the KL condition as shown in Ref.~\cite{knill00}, but knowledge of the initial state of the environment allows for the correction of a potentially smaller operator subspace.

\mysection{General perturbation}
A quantum channel $\chan N$ can always be written as $\chan N(\rho) = \sum_i \lambda_i M_i \rho M_i^\dagger$ where $\lambda_i>0$ and $\tr(M_i^\dagger M_j) = \delta_{ij}$. Indeed a channel can be seen as a positive linear operator, the Choi matrix, whose eigenstates are linearly related to the operators $M_i$, with corresponding eigenvalues $\lambda_i$. Therefore, from linear perturbation theory we know that if $\chan N$ is analytic in a parameter $\tortau$, then so are the operators $M_i$ and the eigenvalues $\lambda_i$. This implies that in general the Kraus operators
\(
N_i = \sqrt{\lambda_i} M_i
\)
are given by power series in $\sqrt{\tortau}$.   
If $\chan N(\rho)=\rho$ for all $\rho$ at $\tortau = 0$ (weak noise), the most general expansion satisfies
\begin{equation}
\chan N(\rho) = \rho + \tortau \chan L(\rho) + \mathcal O(\tortau^2)
\end{equation}
for
\begin{equation}
\chan L(\rho) = -i[H,\rho]  + \sum_{i=1}^m L_i \rho L_i^\dagger - \frac 1 2 \{L_i^\dagger L_i,\rho \}
\end{equation}
and $H^\dagger = H$. This is the familiar term entering the Lindblad equation, although this equation holds also for non-markovian families of channel. It can be obtained by writing $\lambda_i$ and $M_i$ as power series in $\tortau$, with the requirements that the eigenvalue equation holds to zeroth order, and that the channel be normalized. In terms of the Kraus operators expansion (Equ. \ref{exp}) this means that $\epsilon = \sqrt{\tortau}$, $E_0 = 0$ and $E_i = L_i$ for $i=1,\dots,m$. 
Therefore, in this context our result states that necessary and sufficient conditions for correctability to order $\tortau$ are given by the KL conditions applied the all Lindblad operators $L_i$ (and the identity), namely, for all $i$ and $j$,
\begin{equation}
\label{linbdlequ}
P L_i^\dagger L_j P \propto P \quad \text{and} \quad P L_i P \propto P.
\end{equation}

\mysection{Weak interaction}
Our result also applies to a model where the noise is caused by the system interacting with its environment via a Hamiltonian $H$ which, for finite-dimensional systems, has the form
\begin{equation}
H = \lambda \sum_{i=1}^n J_i \otimes K_i
\end{equation}
where $J_i$ and $K_i$ are Hermitian, and the operators $J_i$ act on the system while the $K_i$'s act on the environment causing the noise. The parameter $\lambda > 0$ sets the overall strength of the interaction.

If, in addition, we specify a certain initial state for the environment, we obtain an expression of the form given by Equ.~\ref{exp} for the Kraus operators of the resulting channel. The terms of order $\epsilon = t \lambda$ are
\begin{equation}
E_i = \imath \sum_{j=1}^n J_j \bra i K_j \ket 0
\end{equation}
for $i=0,\dots,m$,
where $\imath^2=-1$ and the states $\ket i$ form an orthonormal basis of the environment, including the initial state $\ket 0$. If the initial state of the environment is mixed: $\rho = \sum_i p_i \proj {\psi_i}$, then one can use a purification $\ket 0 = \sum_i \sqrt {p_i} \ket {\psi_i} \otimes \ket {\psi_i}$ and replace $K_j$ by $K_j \otimes \one$. 

In this context it was shown in Ref.~\cite{knill00} that for lowest-order correction it is sufficient to apply the KL conditions to the space $\mathcal J_1 = {\rm Span}(\one,J_1,\dots,J_n)$, which does not require the knowledge of the initial state of the environment for its definition. Here we see that knowledge of the state $\ket 0$ permits to identify the subspace ${\rm Span}(\one,E_1,\dots,E_m)\subseteq \mathcal J_1$ whose satisfaction of the KL conditions is both necessary and sufficient.
In some cases where this subspace is smaller, this allows for better codes. An example will be given below.

\mysection{Main result}
We want to correct (reverse) the channel $\chan N$ on a code represented by the projector $P$, but we do not expect to be able to do it exactly. In order to quantify our success, we need a measure of how close two channels $\chan N$ and $\chan M$ are. We consider two such measures. For any state $\rho$ one can define the entanglement fidelity
\begin{equation}
F_\rho(\chan N,\chan M) := f((\chan N \otimes \id)(\psi_\rho), (\chan M \otimes \id)(\psi_\rho) )
\end{equation}
where $\ket \psi_\rho$ is any purification of $\rho$, $\psi_\rho \equiv \proj{\psi_\rho}$, and
\begin{equation}
\label{fidelity}
f(\rho,\sigma)= \tr \sqrt{ \rho^{\frac 1 2}\, \sigma \rho^{\frac 1 2}} = \max_{U}|\bra{ \psi_\rho} \,\one \otimes U\, \ket{ \psi_\sigma}|
\end{equation}
is the fidelity between two states~\cite{uhlmann76}. Note that some authors rather define the fidelity as the square of this quantity.
Also, $F_\rho(\chan N,\id)^2$, where $\id$ is the identity channel, is Schumacher's entanglement fidelity of $\chan N$~\cite{schumacher96}.

If the support of $\rho$ is $P$, then $F_\rho(\chan N,\chan M)=1$ if and only if $\chan N(\sigma) = \chan M(\sigma)$ for all $\sigma$ supported inside $P$. 
 
We will also consider the worst-case fidelity, which does not depend on an input state $\rho$ and is defined by
\begin{equation}
F_P^{\rm min}(\chan N,\chan M) := \min_{\rho = \rho P} F_\rho(\chan N,\chan M)
\end{equation}
where the minimum is over all states $\rho$ supported on $P$. 

The degree to which a code $P$ is correctable can be measured by
\begin{equation}
\alpha_\rho = \max_{\chan R} F_\rho(\chan R \chan N,\id) 
\end{equation}
where $P$ projects on the support of $\rho$, or alternatively by
\begin{equation}
\alpha_P^{\rm min} = \max_{\chan R} F_P^{\rm min}(\chan R \chan N,\id).
\end{equation}

Clearly for $\epsilon = 0$, $\chan N = \id$ and both of these are equal to $1$. We want to determine what their expansions in $\epsilon$ are, and under what condition the first non-constant and non-zero term can be made to vanish. Note that although $F_\rho(\chan R \chan N,\id)$ is linear in $\chan R \chan N$, the optimal correction channel $\chan R$ may also depend on $\epsilon$, making the whole expression non-linear in $\epsilon$.

We find that the lowest non-constant order of both $\alpha_\rho$ and $\alpha_P^{\rm min}$ is $\epsilon^2$, and that, in both cases, the terms of order $\epsilon^2$ vanish if and only if there exists $\lambda_{ij} \in \mathbb C$ such that
\begin{equation}
\label{zecond}
P E_i^\dagger E_jP = \lambda_{ij} P \quad \text{and} \quad P E_i P = \lambda_{0i} P
\end{equation}
for all $i \neq 0$ and $j\neq 0$.

This result is simpler than what one might have expected, for instance considering Ref.~\cite{leung97}. In this work the authors attempted to find a modification of the KL conditions on the Kraus operators $N_i$. 
One can easily check that our conditions are equivalent to requesting the existence of $\lambda_{ij} \in \mathbb C$ such that, for all $i,j \neq 0$, 
\begin{align}
P N_0^\dagger N_i P &= \lambda_{0i} \epsilon P + \mathcal O(\epsilon^2)\\
P N_i^\dagger N_j P &= \lambda_{ij} \epsilon^2 P + \mathcal O(\epsilon^3).
\end{align}

\mysection{Independent errors}
Consider $n$ independent copies of the channel $\chan N$ acting in parallel, then the overall channel on those $n$ systems has Kraus operators
\begin{align}
M_0 &= \one + \epsilon \sum_{i=1}^n E_0^{(i)} + \mathcal O(\epsilon^2)\\
M_{ij} &= \epsilon E_j^{(i)} + \mathcal O(\epsilon^2)
\end{align}
where $j \neq 0$, and we write $X^{(i)}$ for the operator $X$ acting on the $i$th system. The other Kraus operators have order $\epsilon^2$ and can be neglected for lowest order correction. 
We see that all the operators $E_0^{(i)}$ can be ignored since they only appear in the Kraus operator $M_0$ which contains the identity component.

For instance, we suppose that the noise on each qubit is due to an interaction of the form $H = J_0 \otimes K_0 + J_1 \otimes K_1$ with $[J_0,J_1] \neq 0$. With no knowledge of the initial state of the environment, one is forced to use a quantum code, which may require 5 physical qubit to encode just one qubit. 

However, with knowledge of the environment's initial state (say $\ket 0$), and our present result, we see that only $E_1 = i\sum_j J_j \bra 1 K_j \ket 0$ matters. If furthermore $E_1$ is normal (which happens here whenever $\real \bra 1 K_1 \ket 0\imag \bra 1 K_0 \ket 0 = \real \bra 1 K_0 \ket 0\imag \bra 1 K_1 \ket 0 $), then the linear span of $\one$ and $E_1$ is equal to the span of $\one$ and $\sigma_z$ defined in the eigenbasis of $E_1$. This implies that a simple repetition code can be used, necessitating only 3 physical qubits to encode one logical qubit, or more generally any classical code correcting one error~\cite{braunstein96}. 

\mysection{Amplitude damping channel}
We can readily apply our result to the amplitude damping channel on $n$ qubits considered in Ref.~\cite{leung97}. The channel on each qubit has Kraus operators
\begin{equation}
N_0 = \one - \mathcal O(\epsilon^2) \quad \text{and}\quad N_1 = \epsilon \ketbra 0 1.
\end{equation}
It describes the decay of a particle from an excited state $\ket 1$ to a ground state $\ket 0$ with probability proportional to $\epsilon^2$. 
Given the analysis of the previous section we see that it is sufficient and necessary to correct the errors $\ketbra 0 1$ acting on any one qubit. However we have no simple way of reducing this to one Pauli error as $\ketbra 0 1$ is not normal.

\mysection{Proof of the main result}
It was shown in Ref.~\cite{beny10} that the quantity
\begin{equation}
\label{delta1}
\Delta_{\rho,\sigma} := \tr \sqrt{\sum_{ij} N_i \rho^2 N_j^\dagger \tr(\sigma N_i^\dagger N_j)}
\end{equation}
provides a good estimate for $\alpha^{\min}_P$ in the sense that
\begin{equation}
\label{ineq1}
\min_{\rho = \rho P} \Delta_{\rho,\sigma} \le \alpha^{\min}_P \le \frac 3 4\min _{\rho = \rho P} \Delta_{\rho,\sigma} + \frac 1 4
\end{equation}
for any state $\sigma = \sigma P$. 

This object $\Delta_{\rho,\sigma}$ measures how much information the ``environment'' obtains about the code. Indeed we can write (as in Ref.~\cite{beny10})
\begin{equation}
\label{delta2}
\Delta_{\rho,\sigma} = F_\rho(\widehat{\chan N},\widehat{\chan N}(\sigma) \tr)
\end{equation}
where $\widehat{\chan N}(\sigma) \tr$ denotes the constant channel with output $\widehat{\chan N}(\sigma)$. 
The hats denote {\em complementary} channels; any channel
\(
\chan M(\rho) = \sum_i M_i \rho M_i^\dagger
\)
can be written as resulting from some unitary interaction with an extra system: the environment. The information that the environment then receives about the initial state is given by the complementary channel
\begin{equation}
\widehat {\chan M}(\rho) := \sum_{ij} \tr(M_i \rho M_j^\dagger) \ketbra i j.
\end{equation}
Hence $\Delta_{\rho,\sigma}$ measures how close $\widehat{\chan N}$ is from a constant channel, i.e. a channel carrying no information about its source. 

The fact that the expressions in Equ.~\ref{delta1} and Equ.~\ref{delta2} are equal can be checked by using the definition of the fidelity as the right hand side of Equ.~\ref{fidelity}, and noting that it has the form $\max_U|\tr(XU)|$ for some operator $X$, which is reached when $U$ is given by the polar decomposition of $X$, yielding $\max_U|\tr(XU)|=\tr(\sqrt{X^\dagger X})$.

Equ.~\ref{ineq1} implies that for any $\nu>0$ and any given $\sigma$ inside the code,
 \(
\Delta_{\rho,\sigma} = 1 + o(\epsilon^\nu) 
\)
for all $\rho$ in the code
if and only if 
\(
\alpha^{\min}_P = 1 + o(\epsilon^\nu)
\)
.
 
Similarly, it was proven in Ref.~\cite{tyson09x1} that
\begin{equation}
\Delta_{\rho,\rho} \le \alpha_\rho \le \sqrt{\Delta_{\rho,\rho}}.
\end{equation}
Hence $\alpha_\rho = 1 + o(\epsilon^\nu)$ if and only if 
\(
\Delta_{\rho,\rho} = 1 + o(\epsilon^\nu)
\)

Our strategy is to compute the first nontrivial order of $\Delta_{\rho,\sigma}$ and find conditions under which this order vanishes in the case $\rho = \sigma$. This solves the problem concerning $\alpha_\rho$. For $\alpha_P$, we will just note that the same conditions are clearly necessary for that same order to vanish, but also sufficient, as a direct calculation shows they make $\Delta_{\rho,\sigma}$ independent from $\sigma$ to that order.

We use degenerate perturbation theory to find the eigenvalues of the operator $\sum_{ij} N_i \rho^2 N_j^\dagger \tr(\sigma N_i^\dagger N_j)$ up to order $\epsilon^{4}$ (see Appendix~\ref{appproof}). Expanding the square root of each eigenvalues in powers of $\epsilon$ and summing them we obtain
\begin{equation}
\begin{split}
\Delta_{\rho,\sigma} = 1 &+ \epsilon^2 \sum_{j\neq0} \re[\tr (\rho E_j^\dagger)\tr (\sigma E_j)] \\
&- \epsilon^2 \sum_{i\neq0}\frac 1 2 \tr[(\rho+\sigma) E_i^\dagger E_i] \\
&+ \epsilon^2 \,\tr\sqrt{\sum_{ij\neq0}P^\perp E_i\rho^2 E_j^\dagger P^\perp \tr[\sigma (E_i')^\dagger E_j']} + o(\epsilon^2)\\
\end{split}
\end{equation}
where $P^\perp = \one - P$ projects on the kernel of $\rho$, and
\begin{equation}
E_i' := E_i - \tr(\rho E_i) \one.
\end{equation}
We first consider the case $\rho = \sigma$ as explained above.
Let us define the non-trace-preserving (nor even trace-decreasing) completely positive maps
\(
\chan E(\rho) = \sum_{i\neq0} E_i \rho E_i^\dagger
\)
and
\(
{\chan E}'(\rho) = \sum_{i\neq0} E_i' \rho (E_i')^\dagger.
\)
Using these definitions we can write
\begin{equation}
\Delta_{\rho,\rho} = 1 - \epsilon^2 \tr\chan E'(\rho) + \epsilon^2 F_\rho(\widehat{\chan P^\perp \chan E}, \widehat{\chan E}'(\rho) \tr) + \mathcal O(\epsilon^3)
\end{equation}
where $\chan P^\perp(\rho) = P^\perp \rho P^\perp$.
The expression for the third term is obtained formally in the same way that Equ.~\ref{delta2} is obtained from Equ.~\ref{delta1}.

Therefore the conditions for the first order term in the fidelities to vanish is
\begin{equation}
\label{baseequ}
\tr\chan E'(\rho) = F_\rho(\widehat{\chan P^\perp \chan E}, \widehat{\chan E}'(\rho) \tr).
\end{equation}
Since the fidelity above is just the overlap between two states, it is bounded by the product of the norm of these states:
\begin{equation}
F_\rho(\widehat{\chan P^\perp \chan E}, \widehat{\chan E}'(\rho) \tr)^2 \le {\tr [P^\perp \chan E(\rho)] \tr [\chan E'(\rho)]}.
\end{equation}
Hence for Equ.~\ref{baseequ} to be satisfied we must have
\begin{equation}
\tr \chan E'(\rho) \le \tr P^\perp \chan E(\rho).
\end{equation}
Direct calculations shows that this means
\begin{equation}
\tr(P\chan E'(\rho)) = \tr(P\chan E'(\rho) P) \le 0.
\end{equation}
But then $P\chan E'(\rho) P = 0$ because it is a positive operator. Indeed, it is a sum of manifestly positive operators of the form $X X^\dagger$:
\begin{equation}
P\chan E'(\rho) P = \sum_i (PE_i'\sqrt \rho)(PE_i'\sqrt \rho)^\dagger = 0.
\end{equation}
This in turns implies that each of the operators $PE_i'\sqrt \rho$ vanishes:
\begin{equation}
PE_i'\sqrt \rho = 0
\end{equation}
for all $i \neq 0$, or simply, recalling that $P$ projects on the support of $\rho$,
\begin{equation}
PE_iP \propto P
\end{equation}
for all $i \neq 0$.

Together with Equ.~\ref{baseequ}, this also implies that
\begin{equation}
F_\rho(\widehat{\chan P^\perp \chan E}, \widehat{\chan E}'(\rho) \tr)^2 = \tr [P^\perp \chan E(\rho)]\, \tr [\chan E'(\rho)]
\end{equation}
Hence the fidelity is maximal. Since also both the states that the fidelity compares have the same norm, they must be equal. This implies that both CP maps are actually equal, namely
\begin{equation}
\widehat{\chan P^\perp \chan E}(\sigma) = \widehat{\chan E}'(\rho) \tr(\sigma)
\end{equation}
for all states $\sigma$ supported on $P$. 
This means that for all $i,j\neq0$,
\begin{equation}
\tr(P^\perp E_i \sigma E_j^\dagger) = \tr[E_i' \rho (E_j')^{\dagger}] \tr(\sigma)
\end{equation}
which is equivalent to 
\begin{equation}
\tr(P E_j^\dagger P^\perp E_i P \sigma ) = \tr(E_i' \rho (E_j')^\dagger) \tr(P \sigma)
\end{equation}
being true for {\it any} state $\sigma$, which in turn implies the operator equation 
\begin{equation}
P E_j^\dagger P^\perp E_i P =  \tr[E_i' \rho (E_j')^\dagger]  P
\end{equation}
for all $i \neq 0$ and $j \neq 0$. Since $PE_iP \propto P$ this also implies
\begin{equation}
P E_i^\dagger E_j P \propto P.
\end{equation}
Hence we have shown the necessity of the conditions expressed in Equ.~\ref{zecond}. 
The sufficiency is straightforward, as both conditions together can be easily checked to imply the equality of $\widehat{\chan P^\perp \chan E}$ and $\widehat{\chan E}'(\rho)$.


\mysection{Acknowledgments}
The author would like to thank Markus Grassl, Daniel Gottesmann, Prabha Mandayam, Milan Mosonyi, Hui Khoon Ng and Ognyan Oreshkov for discussions about this work. This work was supported in part by the cluster of excellence EXC 201 “Quantum Engineering and Space-Time Research”.
We also acknowledge the support by the EU projects CORNER and COQUIT.
The Centre for Quantum Technologies is funded by the Singapore Ministry of Education and the National Research Foundation as part of the Research Centres of Excellence programme.

\bibliography{perturbative_qec}

\newpage
\appendix

\section{Perturbation theory}
\label{appproof}

We suppose that
\begin{equation}
N_i = \delta_{i0} \one + \epsilon E_i + \epsilon^2 F_i + \epsilon^3 G_i + \dots.
\end{equation}
For the channel $\chan N(\rho) = \sum_i N_i \rho N_i$ to be trace-preserving we need in particular that $E_0^\dagger = -E_0$. 
Let us write the operator 
\begin{equation}
S = \sum_{ij} N_i \rho^2 N_j^\dagger \tr(\sigma N_i^\dagger N_j)
\end{equation}
as a power series in $\epsilon$:
\begin{equation}
S = S_0 + \epsilon S_1 + \epsilon^2 S_2 + \epsilon^3 S_3 + \epsilon^4 S_4 + \mathcal O(\epsilon^5).
\end{equation}
Note that 
\begin{equation}
S_0 = \rho^2 \quad \text{and} \quad S_1 = [E_0,\rho^2]
\end{equation}
It is straighforward but tedious to write the operators $S_i$ in terms of the operators $E_i$, $F_i$ and the next order $G_i$. We will not write them explicitly here. 

Let $\lambda_i$ be the $i$th eigenvalue of $S$. We have
\begin{equation}
\Delta_{\rho,\sigma} = \sum_i \sqrt{\lambda_i}.
\end{equation}
Let us write
\begin{equation}
\lambda_i = \sum_n \epsilon^n \lambda_i^{(n)}.
\end{equation}
with the corresponding eigenstates
\begin{equation}
\ket {\lambda_i} = \sum_n \epsilon^n \ket {n,i}.
\end{equation}
As a reference, we will use the basis
\begin{equation}
\ket i := \ket {0,i}.
\end{equation}

We want the eigenvalue equation
\begin{equation}
S\ket {\lambda_i} = \lambda_i \ket{\lambda_i}
\end{equation}
to be satisfied to all orders in $\epsilon$.
For $\epsilon = 0$ we obtain the equation
\begin{equation}
S_0 \ket{i} = \lambda_i^{(0)} \ket i.
\end{equation}
Hence $\ket{i}$ is an eigenbasis of $S_0$. We assume without loss of generality that $\lambda_i^{(0)} = 0$ for $i > d$ and nonzero for $i \le d$. Let
\begin{equation}
P := \sum_{i\le d} \proj i
\end{equation}
be the projector on the range of $S_0$.

We also introduce the ``propagator''
\begin{equation}
D_i = (\lambda_i^{(0)}\one - S_0)^{-1}
\end{equation}
defined to send the kernel of $\lambda_i^{(0)}\one - S_0$ to zero. 
Also, for $i>d$ we write 
\begin{equation}
D \equiv D_i = - S_0^{-1}. 
\end{equation}

For order $n>0$ and all $i$, we obtain the equations
\begin{equation}
\label{pert1}
(\lambda_i^{(0)}\one - S_0)\ket{n,i} = \sum_{m=1}^n (S_m - \lambda^{(m)}_i\one)\ket{n-m,i}
\end{equation}
Multiplying by $\bra i$, this yields
\begin{equation}
\begin{split}
\lambda^{(n)}_i &= \bra i S_n \ket{i} + \sum_{m=1}^{n-1} \bra i \,S_m - \lambda^{(m)}_i\one\, \ket{n-m,i}\\
\end{split} 
\end{equation}
Hence, in order to find $\lambda^{(n)}_i$ we need to know the states $\ket{m,i}$, $m < n$. 

In order to compute $\lambda_i^{(2)}$, consider the projector $P_i$ on the degenerate eigenspace corresponding to eigenvalue $\lambda_i^{(0)}$, so that
\begin{equation}
P_i \ket i = \ket i.
\end{equation}
From Equ.~\ref{pert1} we have
\begin{equation}
P_i^\perp \ket{n,i} = \sum_{m=1}^n D_i (S_m - \lambda^{(m)}_i\one)\ket{n-m,i}.
\end{equation}
Note that in our case $P_i S_1 P_i = 0$, hence $\bra i S_1 = \bra i S_1 P_i^\perp$. It follows that 
\begin{equation}
\begin{split}
\lambda_i^{(2)} &= \bra i S_2 \ket i + \bra i S_1 - \lambda_i^{(1)} \ket {1,i}\\
&= \bra i S_2 \ket i + \bra i (S_1 - \lambda_i^{(1)}) P_i^\perp \ket {1,i}\\
&= \bra i S_2 \ket i + \bra i S_1 D_i S_1 \ket i\\
\end{split}
\end{equation}
where we used that $\lambda_i^{(1)} = 0$. 

We will see that for $i \le d$ we do not need to go to higher order. Therefore we now focus on the cases $i > d$. For these terms, $\lambda_i^{(2)} = 0$. Indeed, noting that $P_i = P^\perp$, we have
\begin{equation}
P^\perp K_2 P^\perp = 0
\end{equation}
where we defined
\begin{equation}
K_2 := S_1 D S_1 + S_2.
\end{equation}
For the next order, Equ~\ref{pert1} yields 
\begin{equation}
\begin{split}
P\ket{2,i}= D S_1 \ket{1,i} + D S_2 \ket{i}\\
\end{split}
\end{equation}
and
\begin{equation}
\begin{split}
- S_0 \ket{3,i} &= S_1 \ket{2,i} + S_2 \ket{1,i} + (S_3 - \lambda_i^{(3)}) \ket{i} \\
\end{split}
\end{equation}
from which
\begin{equation}
\begin{split}
\lambda_i^{(3)} &= \bra i S_1 P \ket{2,i} + \bra i S_2 \ket{1,i} + \bra i S_3\ket{i} \\
&= \bra i\,  S_1 D S_1 + S_2 \,\ket{1,i} + \bra i\, S_3 + S_1 D S_2 \, \ket{i}\\
&= \bra i\, (S_1 D S_1 + S_2)P \,\ket{1,i} + \bra i\, S_3 + S_1 D S_2 \, \ket{i}\\
&= \bra i\, S_1 D K_2 + S_2 D S_1  + S_3 \, \ket{i}.
\end{split}
\end{equation}
As it turns out, again
\begin{equation}
P^\perp K_3 P^\perp = 0
\end{equation}
where we defined
\begin{equation}
K_3 := S_1 D K_2 + S_2 D S_1  + S_3 = K_2 D S_1 + S_1 D S_2 + S_3
\end{equation}
and therefore $\lambda_i^{(3)} = 0$. 
Using 
\begin{equation}
\begin{split}
P\ket{3,i}= D S_1 \ket{2,i} + D S_2 \ket{1,i} + D S_3 \ket{i} \\
\end{split}
\end{equation}
and
\begin{equation}
\label{pert4}
- S_0 \ket{4,i} = S_1 \ket{3,i} + S_2 \ket{2,i} + S_3 \ket{1,i} + (S_4 - \lambda_i^{(4)}) \ket{i} \\
\end{equation}
we finally obtain 
\begin{equation}
\begin{split}
\lambda_i^{(4)} &= \bra i S_1 P \ket{3,i} + \bra i S_2 \ket{2,i} + \bra i S_3 \ket{1,i} + \bra i S_4 \ket{i} \\
&= \bra i K_2 P \ket{2,i} + \bra i\, S_1 D S_2 + S_3\, \ket{1,i} + \bra i \, S_4 + S_1 D S_3 \,\ket{i}\\
&= \bra i K_3 P \ket{1,i} + \bra i \, S_4 + S_1 D S_3 + K_2 D S_2 \,\ket{i}\\
&= \bra i K_3 D S_1 + K_2 D S_2 + S_1 D S_3 + S_4\,\ket{i}.\\
\end{split}
\end{equation}
This is the first nonzero term. We still need to know more about the state $\ket i$. From Equ.~\ref{pert4},
\begin{equation}
\bra j K_3 D S_1 + K_2 D S_2 + S_1 D S_3 + S_4\,\ket{i} = 0
\end{equation}
for all $i,j >d$ and $i \neq j$. This means that the states $\ket i $ for $i > d$ must be eigenstates of the operators $P^\perp K_4 P^\perp$, where
\begin{equation}
K_4 := K_3 D S_1 + K_2 D S_2 + S_1 D S_3 + S_4
\end{equation}
and $\lambda_i^{(4)}$ are the corresponding eigenvalues. 

Since $S_0 = \rho^2$, we have $\lambda_i^{(0)} = p_i^2$; where $p_i$ are the eigenvalues of $\rho$ and $\ket i$ its eigenstates. 

Putting everything together, we obtain
\begin{equation}
\begin{split}
\Delta_{\rho,\sigma} &= \sum_{i\le d} \sqrt {\lambda_i^{(0)} + \epsilon^2 \lambda_i^{(2)} + \mathcal O(\epsilon^3)} + \epsilon^2 \sum_{i > d} \sqrt{\lambda_i^{(4)}}\\
&= \sum_{i\le d} \left[{ p_i + \frac 1 2 \epsilon^2 \frac {\lambda_i^{(2)}}{p_i}}\right]  + \epsilon^2 \sum_{i > d} \sqrt{\lambda_i^{(4)}} + \mathcal O(\epsilon^3)\\
\end{split}
\end{equation}
where a direct calculation yields
\begin{equation}
\begin{split}
\sum_{i\le d} \frac {\lambda_i^{(2)}}{2p_i} 
&=  \sum_{j>0} \re \tr (\rho E_j^\dagger) \tr(\sigma E_j) - \frac 1 2 \sum_{i>0}\tr((\sigma+\rho) E_i^\dagger E_i). \\
\end{split}
\end{equation}
and
\begin{equation}
\begin{split}
\sum_{i>d}\sqrt{\lambda_i^{(4)}} &= \tr\sqrt{\sum_{ij>0}P^\perp E_i\rho^2 E_j^\dagger P^\perp \tr(\sigma (E_i')^\dagger E_j')}\\
\end{split}
\end{equation}
where $E'_i = E_i - \tr(\sigma E_i)$.

\end{document}